\documentclass[10pt]{article}
\usepackage{hyperref}
\usepackage{float}
\usepackage[symbol]{footmisc}
\usepackage[superscript,sort]{cite}
\usepackage{array}
\usepackage{bm}

\newcolumntype{x}[1]{%
>{\centering\hspace{0pt}}p{#1}}%
\usepackage[normalem]{ulem} 
\usepackage{amsmath,amssymb}

\usepackage{amsthm,amscd,amsxtra,amsfonts,amsmath,amssymb,multirow}
\usepackage{wrapfig}
\usepackage[tiny,compact]{titlesec}
\usepackage{threeparttable} 
\usepackage{helvet}

\usepackage{booktabs}
\usepackage[table]{xcolor}
\usepackage{xcolor,colortbl}
\usepackage{placeins}
\usepackage{tikz}
\usetikzlibrary{shapes,arrows}
\usepackage[T1]{fontenc}
\usepackage[linesnumbered,ruled]{algorithm2e} 
\titlespacing*{\section}{0pt}{*0}{*0}
\titlespacing*{\subsection}{0pt}{*0}{*0}
\titlespacing*{\subsubsection}{0pt}{*0}{*0}
\titlespacing{\paragraph}{0pt}{*0}{*1}

\definecolor{MyPurple}{rgb}{1,0,1}

\newcommand{\beq}[1]{\begin{equation} \label{#1}}
\newcommand{\eeq}{\end{equation}}
\newcommand{\barray}{\begin{array}{ll}}
\newcommand{\earray}{\end{array}}

\usepackage{longtable} 
\usepackage{threeparttable} 

\begin{document}

\title{Algebraic graph learning of protein-ligand binding affinity}
\author{Duc Duy Nguyen$^{1}$ and  Guo-Wei Wei$^{1,2,3,}$\footnote{
        Corresponding to Guo-Wei Wei.       Email: wei@math.msu.edu}\\
$^1$ Department of Mathematics,
Michigan State University, MI 48824, USA.\\
$^2$ Department of Electrical and Computer Engineering,
Michigan State University, MI 48824, USA. \\
$^3$ Department of Biochemistry and Molecular Biology,
Michigan State University, MI 48824, USA. \\
}

\date{\today}
\maketitle

\begin{abstract}
Although algebraic graph theory based models have been  widely applied in physical modeling and molecular studies, they are typically incompetent in the analysis and prediction of biomolecular properties when compared with other quantitative approaches. There is a need to explore the capability and limitation of algebraic graph theory for molecular and biomolecular modeling, analysis, and  prediction. In this work, we propose novel algebraic graph learning (AGL) models that encode  high-dimensional  physical and biological information into intrinsically low-dimensional representations. The proposed AGL model introduces multiscale weighted colored subgraphs to describe  crucial molecular and biomolecular interactions  via graph invariants associated with the graph Laplacian, its pseudo-inverse, and  adjacent matrix.  Additionally, the AGL models are incorporated with an advanced machine learning algorithm to connect the low-dimensional graph representation of biomolecular structures with their macroscopic properties. Three popular protein-ligand binding affinity benchmarks, namely CASF-2007, CASF-2013, and CASF-2016, are employed to validate the accuracy, robustness, and reliability of the present AGL model. Numerical results indicate that the proposed AGL method outperforms the other state-of-the-art methods in the binding affinity predictions of the protein-ligand complexes.
\end{abstract}

\section{Introduction}

Graph theory is a prime subject of discrete mathematics that concerns graphs as mathematical structures for modeling pairwise relations between vertices, nodes, or points. Such pairwise relations define graph edges. There are many different graph theories, such as geometric graph theory, algebraic graph theory, and topological graph theory.  Geometric graphs admit geometric objects as graph nodes or vertices. algebraic graph theory, particularly spectral graph theory, studies the algebraic connectivity via characteristic polynomial, eigenvalues, and eigenvectors of matrices associated with graphs, such as adjacency matrix or Laplacian matrix. Topological graph theory concerns the embeddings and immersions of graphs, and the association of graphs with topological spaces, such as abstract simplicial complexes. Mathematically, graphs are useful tools in geometry and certain parts of topology such as knot theory and algebraic topology.

Like topology, graph theory also emphasizes the connectivity. The geometric connectivity  of a graph   refers to pairwise relations among graph nodes and is often analyzed by ``topological index'' \cite{hosoya1971topological,hansen1988chemical}, contact map \cite{Bahar:1997, LWYang:2008} and  graph centrality \cite{newman2010networks,bavelas1950communication,dekker2005conceptual}. The algebraic connectivity of a graph refers to the second-smallest eigenvalue of the Laplacian matrix of the graph  and  is also known as Fiedler value or Fiedler eigenvalue, which has many applications, including the stability analysis of dynamical systems \cite{Wei:2002e}.
In contrast, topological connectivity refers to the connectedness of the entire system rather than pairwise ones as in the geometric graph theory.  Topological connectivity is an important property for distinguishing topological spaces.

Graph theory has been widely applied in physical, chemical, biological, social, linguistic, computer  and information sciences. Many practical problems can be represented and analyzed by  graphs.  In chemistry and biology,  a graph makes a natural model for a molecule, where graph vertices represent atoms and graph edges represent possible bonds.   Graphs have been widely  used in chemical analysis  \cite{trinajstic1983chemical,schultz1989topological,  janezic2015graph} and biomolecular modeling \cite{Angeleska:2009}, including normal mode analysis (NMA)  \cite{Go:1983,Tasumi:1982,Brooks:1983,Levitt:1985}  and elastic network model (ENM) \cite{Flory:1976, Bahar:1997,Bahar:1998,Atilgan:2001,Hinsen:1998,Tama:2001} 
for modeling  protein flexibility and long-time dynamics. Some of the most  popular ENMs are Gaussian network model (GNM)   \cite{Bahar:1997,Bahar:1998,QCui:2010}  and anisotropic network model (ANM) \cite{Atilgan:2001}. In these methods,  the  diagonalization of the interaction Laplacian matrix is a required procedure to analyze protein flexibility, 
which has the computational complexity of  ${\mathcal O}(N^3)$ with $N$ being the number of matrix elements. Graph theory has also been used to represent the structures of molecules and biomolecules, resulting in a popular approach for  chemical  datasets \cite{balaban1976chemical,trinajstic1983chemical, schultz1989topological, foulds2012graph, hansen1988chemical, ozkanlar2014chemnetworks} and biomolecular datasets \cite{Bahar:1997,di2015protein, canutescu2003graph, ryslik2014graph,Jacobs:2001, vishveshwara2002protein,wu2017moleculenet} in the past few decades.

Graph theories, especially geometric graph theories, are relatively intuitive and easy to use. Indeed, a great portion of graph theory based study in molecular and biomolecular systems was qualitative and descriptive. Despite intensive effort in the past, graph theory based quantitative methods are often not as competitive as other quantitative approaches in the analysis and prediction of biomolecular properties from massive and diverse datasets. For example, graph-signature based prediction of protein stability changes upon mutation \cite{Pires:2014} was not as accurate as some other methods \cite{LQuan:2016,ZXCang:2017a, ZXCang:2017c}. Additionally, the average Pearson correlation coefficients in protein B-factor predictions using spectral graph theory based Gaussian network model (GNM) were less than 0.6 in all of three datasets \cite{JKPark:2013}. These situations may be due to the following reasons. Firstly, most graph theory based models do not distinguish different chemical element types in a molecule or biomolecule, which leads to a severe loss of critical chemical and biological information. Secondly, in many molecular graphs, edges are used to represent covalent bonds, while non-covalent interactions are often ignored, which underrepresents the physical interactions of many biomolecular datasets. Finally, many graph-based models approximate the distance between a pair of atoms by counting the number of covalent bonds between them, which leads to a major error in describing their interaction strength.

In the past few years, we have developed a number of graph theory approaches to address the aforementioned problems. For example, weighted graphs were proposed in terms of flexibility-rigidity index (FRI) to represent graph edges by radial basis functions  \cite{KLXia:2017, Opron:2014,Opron:2015a, DDNguyen:2016b}. Physically, we assume that protein interactions, including those with its environment,  fully determine its structure at the equilibrium. Protein structure and its environment, in turn, fully determine protein flexibility and function. As a consequence, one does not need to invoke a high-dimensional theoretical model that is subject to modeling errors  to analyze protein flexibility and function when the native structure of the protein and its environment is known. Mathematically, our approach assumes a complete graph while weights the importance of graph edges by scaling their Euclidean distances in  radial basis functions so that the nearest neighbors in the sense of the Euclidean metric have the strongest edges. Additionally, multiscale FRI is a multigraph approach which is permitted to have multiple edges \cite{Opron:2015a,KLXia:2015f}. Similar to persistent homology \cite{Edelsbrunner:2002, Zomorodian:2005},  this multi-edge technique allows a given molecular graph to be analyzed in multiscale, capturing the multiscale interactions in macromolecules \cite{Opron:2015a}.  Graph coloring, or more generally, graph labeling, is an important graph theory technique for graph vertices or edges to be treated differently. This method enables the encoding of chemical and biological information into molecular graphs \cite{DDNguyen:2017d,DBramer:2018a}. Subgraphs constructed from vertex-labeled graphs and edge-labeled graphs give rise to powerful graph representations of intermolecular and intramolecular interactions,  such as hydrogen bonds, electrostatics, van der Waals interactions,  hydrophilicity,  hydrophobicity etc  \cite{DDNguyen:2017d,DBramer:2018a}. Our multiscale weighted colored graph is over 40\% more accurate than GNM in protein B-factor predictions. This approach outperformed all other methods in protein-ligand binding affinity predictions \cite{DDNguyen:2017d} and its predictions of free energies and their  rankings (Kendall's tau) were ranked 1st in of Set 1 (Stage 2) of  \href{https://drugdesigndata.org/php/d3r/gc2/combined/free-energy/index.php?component=443&q=rmsd&set=1}{D3R Grand Challenge 2} and in 10 of a total 26 contests in \href{https://drugdesigndata.org/about/grand-challenge-3-evaluation-results}{D3R Grand Challenge 3}.

  Conceptually, our approach is built upon the fundamental hypothesis that intrinsic physics of interest lies in low-dimensional subspaces or manifolds embedded in a high-dimensional data space. While the hypothesis is quite well-known in manifold learning,  the major challenge is how to encode crucial physical information contained in the high-dimensional space into a desirable low-dimensional representation of molecules and/or biomolecules. Our multiscale weighted colored (or labeled) subgraphs address this challenge. A major advantage of our multiscale weighted colored subgraph approach is its low-dimensionality, simplicity, and robustness. For example, only required data inputs are atomic names and coordinates. Indeed,  it bypasses complicated data processing and parametrization. It does not need any quantum mechanical (QM) and molecular mechanical (MM) force fields, namely, charges,  polarization assignments, bond lengths and angles,  van der Waals well depths, dielectric constants, surface tension,   electronegativity, etc. As such, it avoids errors associated with the parametrization. In fact, our graph theory approach is also simpler than our algebraic topology approach mathematically and computationally, while it performs as well as our topological approach \cite{ZXCang:2018a}.

The objective of the present work is to develop multiscale weighted labeled algebraic subgraphs for representing molecules, biomolecules and their interactions.  For a given geometric (sub)graph, there are many ways to construct corresponding algebraic (sub)graphs. Three most commonly used algebraic graphs are graph Laplacian matrix, its pseudo-inverse, and adjacency matrix. The eigenvalues and eigenvectors computed from these matrices can be used to describe molecules, biomolecules and their interactions in many different ways. We examine a few common approaches in this work.

\section{Methods and  Algorithms}\label{Methods}

\subsection{Multiscale weighted labeled geometric subgraphs }
We propose to  develop systematical, scalable, accurate graph theory descriptors of  protein-ligand binding interactions from massive and diverse datasets. However, the proposed method can be applied to other problems   
such as the predictions of toxicity, solubility, solvation, partition coefficient,  mutation-induced protein folding stability change,  and protein-nucleic acid interactions. 
 In the present work, we target pairwise non-covalent interactions in our subgraph theory description.  For a given dataset, we first perform a statistical analysis to identify a set of commonly occurring chemical element types, say ${\mathcal C}=\{{\rm H, C, N, O, S, P, F, Cl, Br, \cdots }\}$.  For a given  molecule or biomolecule in the dataset, let us denote
\begin{equation}
{\mathcal V} = \{(\mathbf{r}_j, \alpha_j)|\mathbf{r}_j\in {\rm I\!R}^3; \alpha_j \in {\mathcal C};  j=1,2,\ldots,N \}
\end{equation}
a subset of $N$ atoms (i.e., subgraph vertices) that are members of ${\mathcal C}$. Note that the $i$th atom  is labeled both by its element type $\alpha_j$ and its position ${\bf r}_j$. The classification of atoms into    chemical element types  is a graph coloring, which is important for encoding different types of interactions and gives rise to a basis for the collective coarse-grained description of the dataset.  We assume that all the pairwise non-covalent interactions  between element types ${\mathcal C}_{k}$ and ${\mathcal C}_{k'}$ in a molecule or molecular complex can be represented by  fast-decay weight functions
\begin{multline}\label{CollInter}
{\mathcal E}= \{\Phi(||\mathbf{r}_i-\mathbf{r}_j||; \eta_{kk'})| \alpha_i  = {\mathcal C}_{k}, \alpha_j  = {\mathcal C}_{k'};  i,j = 1,2,\ldots,N;
||\mathbf{r}_i-\mathbf{r}_j||> r_i+r_j +\sigma \},
\end{multline}
 where $||\mathbf{r}_i-\mathbf{r}_j||$ is the Euclidean distance between the $i^{th}$ and $j^{th}$ atoms, $r_i $ and $ r_j$ are the atomic radii of $i^{th}$ and $j^{th}$ atoms, respectively and $\sigma$ is the mean value of the standard deviations of $r_i $ and $ r_k$  in the dataset. The distance constraint ($||\mathbf{r}_i-\mathbf{r}_j||> r_i+r_j +\sigma $) excludes covalent interactions. Here $\eta_{kk'}$ is a characteristic distance between the atoms,  
 and $\Phi$ is a subgraph weight and is chosen to have the following properties \cite{Opron:2014}:
 \begin{align}
\Phi(||\mathbf{r}_i-\mathbf{r}_j||;\eta_{kk'})&=1, \mbox{ as } ||\mathbf{r}_i-\mathbf{r}_j||\rightarrow 0 \quad {\rm and }\\
\Phi(||\mathbf{r}_i-\mathbf{r}_j||;\eta_{kk'})&=0 \mbox{ as } ||\mathbf{r}_i-\mathbf{r}_j||\rightarrow\infty,
\quad \alpha_i  = {\mathcal C}_{k}, \alpha_j  = {\mathcal C}_{k'}.
 \end{align}
Although most radial basis functions can be used, generalized exponential functions and  generalized Lorentz functions were shown to work very well for biomolecules \cite{Opron:2014}. We, therefore, have a weighted colored subgraph $G({\mathcal V}, {\mathcal E})$.  To construct element-level  collective molecular descriptors, we propose the multiscale weighted colored subgraph rigidity between
$k$th  element type  ${\mathcal C}_{k}$ and  $k'$th element type ${\mathcal C}_{k'}$

\begin{multline}\label{ESRI}
 {\rm RI}^G(\eta_{kk'}) =\sum_i \mu^G_i(\eta_{kk'})=\sum_i\sum_{j } \Phi(||\mathbf{r}_i-\mathbf{r}_j||;\eta_{kk'}),
\quad \alpha_i  = {\mathcal C}_{k}, \alpha_j  = {\mathcal C}_{k'};  ||\mathbf{r}_i-\mathbf{r}_j||> r_i+r_j +\sigma,
\end{multline}
where $\mu^G_i(\eta_{kk'})$ is a geometric subgraph centrality for the $i$th atom, which offers accurate  protein B-factors predictions  \cite{DBramer:2018a}.  The physical interpretation of Eq. (\ref{ESRI}) is straightforward --- the summation over  $\mu^G_i(\eta_{kk'})$  in Eq. (\ref{ESRI}) leads to the total interaction strength for the selected pair of element types  ${\mathcal C}_{k}$ and ${\mathcal C}_{k'}$, which provides the element-level coarse-grained description of molecular level properties. The above formulation is a generalization of the successful  bipartite subgraph  used in our earlier predictions of protein-ligand binding affinities  and free energy ranking  \cite{DDNguyen:2017d}.
For a bipartite  subgraph,  each of its edges connects one atom in the protein and another atom in the ligand. The graph coloring, i.e., element specific descriptions, and subgraph weight are designed to capture  hydrogen bonds, polarization,  electrostatics, van der Waals interactions,  hydrophilicity,  hydrophobicity, etc.

The different selections  of characteristic distance $\eta_{kk'}$ give rise to a multiscale description of intermolecular and intramolecular interactions.  By appropriate selections of element combinations $k$ and $k'$, the   characteristic distance $\eta_{kk'}$ , and subgraph weight $\Phi$, we systematically construct a family of collective, scalable, multiscale graph-based molecular and biomolecular descriptors. 
\begin{figure}[!htpb]
\centerline{\includegraphics[keepaspectratio,width=3.3in]{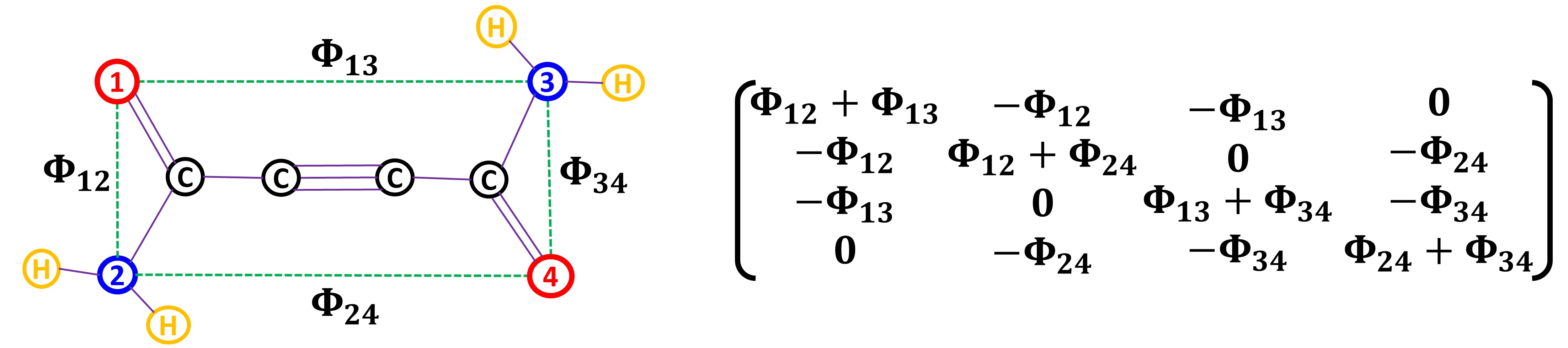}}
\caption{Illustration of   weighted colored subgraph $G_{\rm NO}$ (Left) and its  Laplacian matrix (Right) for cellocidin molecule (C$_4$H$_4$N$_2$O$_2$).  Graph vertices, namely oxygen (i.e., atoms 1 and 4)  and nitrogen (i.e., atoms 2 and 3), are labeled in red and blue colors, respectively. Here,  graph edges (i.e., $\Phi_{ij}$) are labeled by green-dashed lines  which are {\it not} covalent bonds. Note that there are 9 other nontrivial subgraphs for this molecule (i.e., $G_{\rm CC}, G_{\rm CN}, G_{\rm CO}, G_{\rm CH}, G_{\rm NN}, G_{\rm NH}, G_{\rm OO}, G_{\rm OH}, G_{\rm HH}$).
}
\label{fig:Graph}
\end{figure}
The proposed  multiscale weighted colored subgraph rigidity is  simple and robust --- the only required data input is atomic names and coordinates. Indeed,  it bypasses complicated data  processing,  parametrization, and molecular mechanical and quantum mechanical force fields,  such as charges, high-order polarizations, van der Waals well depths, dielectric constants, surface tensions, and  electronegativity,   and their associated errors in many physical models.
Consequently, our graph theory approaches are very fast \cite{Opron:2014}.  Our fast algorithm  has the computational complexity of O($N$) and is able to  predict B-factors for $\alpha$-carbons of an HIV virus capsid (313 236 residues) in less than 30 seconds on a single processor \cite{Opron:2014}.

\subsection{Multiscale weighted labeled algebraic subgraphs }

Our earlier work has demonstrated how to construct powerful geometric graph descriptors for analyzing and predicting biomolecular datasets. Mathematically, it is extremely interesting to understand whether there exist equally powerful algebraic subgraph formulations. Biologically, it is important to develop alternative graph tools for describing molecules, biomolecules and their interactions since each method has its own advantages and potentials.   For a given subgraph, its matrix representations provide a straightforward description of the interaction between subgraph elements, which can be easily expressed by  matrices.   Two most important matrices are the Laplacian matrix and the adjacency matrix. 
\paragraph{Multiscale weighted labeled Laplacian matrix }
We consider a subgraph $G_{kk'} $ for each pair of element types, ${\mathcal C}_k$ and ${\mathcal C}_{k'}$, and
propose an element-level weighted labeled Laplacian matrix $L(\eta_{kk'})$ with elements

\begin{equation} \label{Laplacianmatrix}
L_{ij}(\eta_{kk'}) = \left\{ \begin{array}{ll}
     - \Phi(||\mathbf{r}_i - \mathbf{r}_j||; \eta_{kk'}) &
     \begin{aligned}
        &\text{if}~  i\neq j, \alpha_i  = {\mathcal C}_{k}, \alpha_j  = {\mathcal C}_{k'}\\
        &\text{and}~  ||\mathbf{r}_i-\mathbf{r}_j||> r_i+r_j +\sigma ;
     \end{aligned}
        \\
     -\sum_j L_{ij}  &  {\rm if}~~ i=j .
        \end{array} \right.
\end{equation}

Mathematically, our element-level  weighted labeled Laplacian matrix is  symmetric, diagonally dominant and positive-semidefinite, and thus all of its eigenvalues are nonnegative. Since every row sum or column sum of $L(\eta_{kk'})$ is zero, the first eigenvalue value is zero. The second smallest eigenvalue of $L(\eta_{kk'})$ is the so called algebraic connectivity (or Fiedler value) of $G_{kk'}$. It is interesting to note that
$${\rm RI}^G(\eta_{kk'})={\rm Tr} L(\eta_{kk'}),$$
where ${\rm Tr}$ is the trace. Denote $\lambda^L_j, j=1,2,\cdots$ and ${\bf u}^L_j, j=1,2,\cdots$ the eigenvalues and eigenvectors of  $L(\eta_{kk'})$. We define an atomic descriptor for the $i$th atom $({\bf r}_i, \alpha_i = {\mathcal C}_k)$:
\begin{equation}\label{eq:atomic}
\mu^L_i(\eta_{kk'})=\sum_l (\lambda^L_l)^{-1}\left[{\bf u}^L_l({\bf u}^L_l)^T \right]_{ii},
\end{equation}
 where $T$ is the transpose.  We further propose  element-level weighted labeled Laplacian matrix based molecular descriptors
\begin{equation}
{\rm RI}^L(\eta_{kk'}) =\sum_i \mu^L_i(\eta_{kk'}).
\end{equation}
Note that $\mu^L_i(\eta_{kk'})$ is a   weight subgraph generalization of   GNM \cite{Bahar:1997} or a subgraph generalization of our earlier generalized multiscale FRI \cite{KLXia:2015f}. Therefore,  $\mu^L_i(\eta_{kk'})$ can be used to represent atomic properties, such as  protein B-factors. Additionally,  we can construct a set of element-level weighted labeled Laplacian matrix based molecular descriptors by the statistics of  $\mu^L_i(\eta_{kk'})$, i.e., sum, mean, maximum, minimum and standard deviation, etc., of $\mu^L_i(\eta_{kk'})$.

Alternatively, we can directly construct another set of element-level weighted labeled Laplacian matrix based molecular descriptors by the statistics of  nontrivial eigenvalues $\{\lambda^L_j\}_{j=2,3,\cdots}$. In this case, the Fiedler value is included as the minimum.  The performances of these two sets of molecular descriptors based constructed from  element-level weighted labeled subgraph Laplacian matrix  will be examined and compared.


\paragraph{Multiscale weighted labeled  adjacency matrix }
The element-level weighted labeled adjacency matrix is equally important and can be easily constructed for  subgraph $G_{kk'}$by
\begin{equation} \label{adjacencymatrix}
A_{ij}(\eta_{kk'}) = \left\{ \begin{array}{ll}
     \Phi(||\mathbf{r}_i - \mathbf{r}_j||; \eta_{kk'}) &
    \begin{aligned}
        &\text{if}~  i\neq j, \alpha_i  = {\mathcal C}_{k}, \alpha_j  = {\mathcal C}_{k'}\\
        &\text{and}~  ||\mathbf{r}_i-\mathbf{r}_j||> r_i+r_j +\sigma ;
    \end{aligned}
        \\
   0 &  {\rm if}~~ i=j .
        \end{array} \right.
\end{equation}
Mathematically,  adjacency matrix $A(\eta_{kk'})$  is a  symmetric non-negative matrix  and it contains  the same amount of information as  the corresponding Laplacian matrix, although its eigenvalues
$\lambda^A_j, j=1,2,\cdots$ and  eigenvectors  ${\bf u}^A_j, j=1,2,\cdots$ behave very differently from those of   corresponding Laplacian matrix. The spectrum of the proposed  element-level weighted colored adjacency matrix is real. For each eigenvalue,  its opposite is also an eigenvalue.  Therefore, only positive eigenvalues will be used in our description. The Perron-Frobenius theorem states that the greatest eigenvalue, i.e., the spectral radius $\rho(A)$,  is bounded above by the maximal diagonal element of the corresponding  Laplacian matrix $\min_i\sum_j A_{ij} \leq \rho(A)\leq \max_{i} \sum_j A_{ij}$.  The values of Laplacian matrix elements depend on the scale parameter  $\eta_{kk'}$ and have many zeros at a characteristic scale parameter for hydrogen bonds or van der Waals interactions. However  characteristic scale for electrostatic and hydrophobic interactions can be very large \cite{ZXCang:2017b}. In such as case,   spectral radius $ \max_{i} \sum_j A_{ij}\approx n-1$, with $n$ being the number of atoms in the subgraph $G_{kk'}$.

Assume that all eigenvalues and eigenvectors  of $A_{ij}(\eta_{kk'})$ are given by  $\{\lambda^A_j\}$ and  $\{{\bf u}^A_j\}$, respectively.  In the present work, we use a set  of statistical values, namely,  the sum  of all positive eigenvalues, the mean of all positive eigenvalues, the largest (i.e., the principal) and the smallest positive eigenvalues, and the standard standard deviation of all positive eigenvalues
as  element-level molecular  descriptors of molecules, biomolecules and their interactions.

In principle, we can also construct   atomic descriptors from    $\{\lambda^A_j\}$ and  $\{{\bf u}^A_j\}$. Let us define a square matrix $Q$ whose columns are the $n$ linearly independent eigenvectors of $A$: $Q=[{\bf u}^A_1  {\bf u}^A_2\cdots {\bf u}^A_n]$ and  a diagonal matrix $\Lambda $ where each diagonal element $\Lambda_{ii}$ is the eigenvalue associated with the $i$th column of $Q$. Then, a set of atomic descriptors can be obtained as
\begin{equation}\label{eq:atomicA}
\mu^A_i(\eta_{kk'})= \sum_ j  \left[Q \Lambda Q^{-1}\right]_{ij}.
\end{equation}
{However, the method given in Eq. (\ref{eq:atomicA}) is not a computationally efficient approach for describing atoms in molecules.}


\subsection{Graph learning}

\begin{figure}[!ht]
    \centering
     \includegraphics[width=0.8\textwidth]{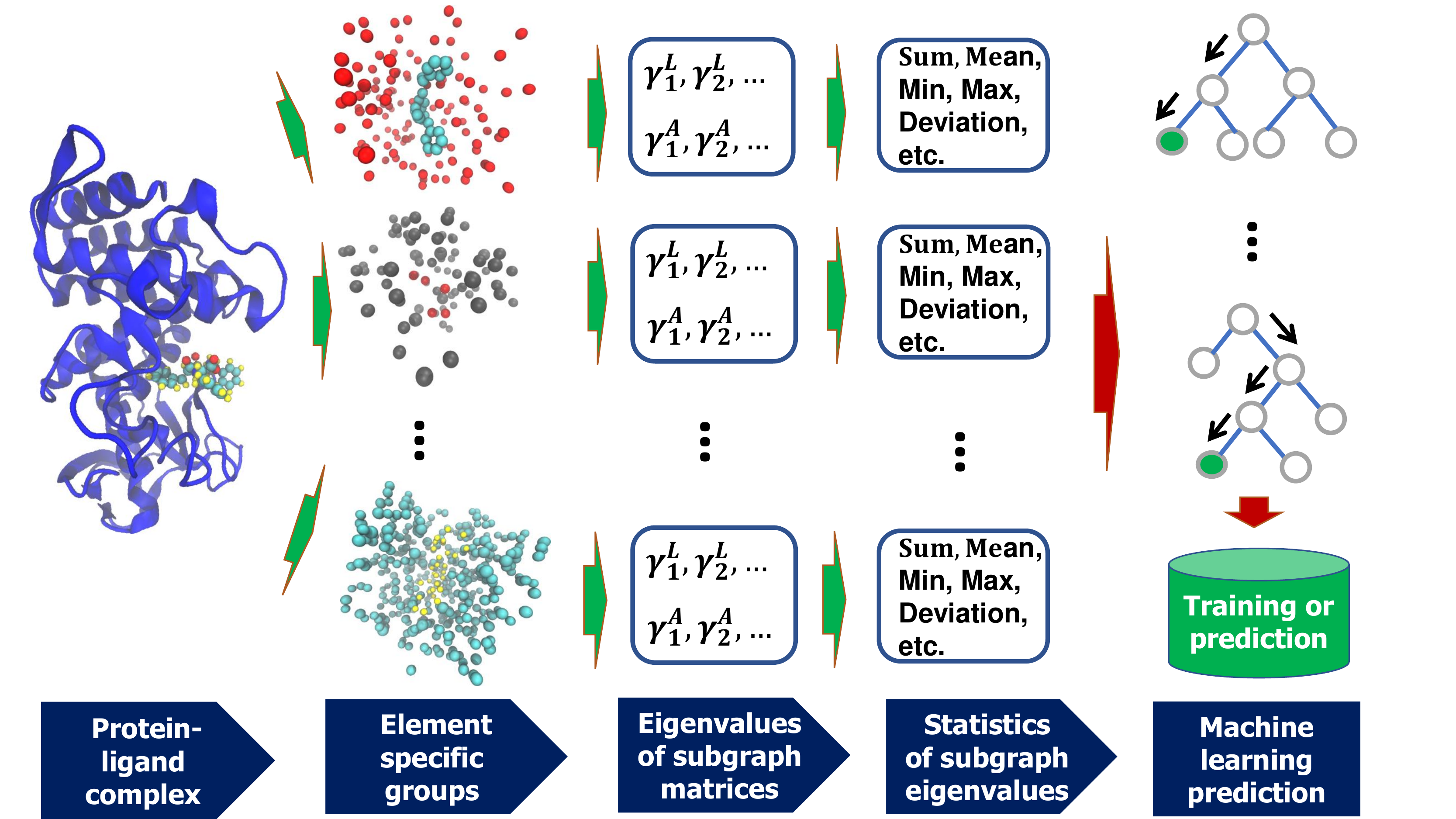}
    \caption{ Illustration of algebraic graph learning strategy using 1OS0 (first column).  In the second column, element specific groups are, from top to bottom, OC, NO, and CH, respectively.  Their corresponding weighted labeled graph Laplacian and adjacency eigenvalues  are shown in the third column. The statistics of these eigenvalues (fourth column) are used in gradient boosting trees for training and prediction (last column).}
    \label{fig:flowchart}
\end{figure}
 To predict molecular and biomolecular properties, statistics of eigenvalues  generated from the proposed weighted labeled subgraph Laplacian matrix or adjacency matrix will be combined with a machine learning algorithm. We assume the dataset is labeled and the problem is either a classification or a regression. From the machine learning point of view, we employ a supervised learning algorithm involving a training set and a test set.  Denote ${\mathcal X}_i$  the dataset from the $i$th molecule or molecular complex in the training dataset and let $\mathbf{G}({\mathcal X}_i; \bm{\zeta})$ be a function that maps the geometric information  into suitable graph representations with a set of parameters $\bm{\zeta}$ consisting of kernel parameters.     To set up a machine learning model, we cast the training into  a minimization problem,
\begin{equation}
\min\limits_{\bm{\zeta}, \bm{\theta}} \sum\limits_{i\in I}{\mathcal L}(\mathbf{y}_i, \mathbf{G}({\mathcal X}_i; \bm{\zeta}); \bm{\theta}),
\end{equation}
where ${\mathcal L}$ is a scalar loss function to be minimized and  $\mathbf{y}_i$ is the collection of labels in the training set. Here $\bm{\theta}$ are the set of machine learning  parameters to be optimized and depend on machine learning algorithms chosen. The loss function ${\mathcal L}$ can be chosen according to the nature of the problem, i.e., regression and classification.   Many  machine learning algorithms, such as random forest, gradient boosting trees, artificial  neural networks, and convolutional neural networks, can be employed in conjugation with the present graph descriptors. However, as our goal in the present work is to examine the descriptive   power of the proposed algebraic graph features,  let us focus on a relatively simple while still powerful machine learning algorithm, gradient boosting trees (GBTs).  GBTs are very robust against overfitting \cite{ZXCang:2017a} and their performance is quite similar to that of random forest.  Figure \ref{fig:flowchart} illustrates  the proposed graph learning strategy.

Throughout this work, we choose  GradientBoostingRegressor module implemented in the scikit-learn v0.19.1 package with  parameters n\_estimators=10000, max\_depth=7, min\_samples\_split=3, learning\_rate=0.01, loss=ls, subsample=0.3, and max\_features=sqrt.  The change in these parameters does not significantly affect the prediction results.

\section{Results}
Herein we assess the scoring power of the proposed algebraic graph learning (AGL) approach on the assay of small molecules and biomolecules. The datasets involve the free binding energy prediction of the interaction between ligand and protein in a complex.

\subsection{Model parametrization}
For the sake of convenience, we use notation ${\rm AGL}^{\mathcal{M}}_{\Omega, \beta, \tau}$ to indicate the algebraic graph learning features generated by using interactive matrix type $\mathcal{M}$ with kernel type  $\Omega$ and corresponding kernel parameters $\beta$ and $\tau$. As such,  $\mathcal{M}={\rm Adj}$, $\mathcal{M}={\rm Lap}$, and $\mathcal{M}={\rm Inv}$ represent adjacent matrix, Laplacian matrix, and pseudo inverse of Laplacian matrix, respectively. Here, $\Omega={\rm E}$ and $\Omega={\rm L}$ refer to generalized exponential and generalized Lorentz kernels, respectively. Additionally, $\beta$ is the kernel order such that $\beta=\kappa$ if $\Omega={\rm E}$, and $\beta=\nu$ if $\Omega={\rm L}$. Finally,  $\tau$ is used such that $\eta_{kk'}=\tau (\bar{r}_k + \bar{r}_{k'})$, where $\bar{r}_k$ and $\bar{r}_{k'}$ are the van der Waals radii of element type  $k$ and element type $k'$, respectively.

We propose an  AGL  representation in which multiple kernels are parametrized at different scale ($\eta$) values. In this work, we consider at most two kernels.  As a straightforward notation extension, two kernels can be parametrized  by  ${\rm AGL}^{\mathcal{M}_1\mathcal{M}_2}_{\Omega_1, \beta_1,\tau_1;\Omega_2,\beta_2,\tau_2}$.

\subsection{Datasets}
In this work, we validate our proposed model against three commonly drug-discovery related benchmarks, namely,
CASF-2007 \cite{RenxiaoWang:2009Compare},  CASF-2013 \cite{YLi:2014}, and  CASF-2016 \cite{su2018comparative}. These benchmarks are collected in the PDBbind  database and have been used  to test the scoring power of various scoring functions (SFs) for  protein-ligand  binding affinities. To train our AGL models, we make use of  refined sets in the PDBbind database, namely, PDBbind v2007 \cite{RenxiaoWang:2009Compare}, PDBbind v2015 \cite{PDBBind:2015}, and PDBbind v2016  \cite{su2018comparative},  for each specific benchmark. The statistical information of these datasets is provided in Table \ref{tab:PDBbind_datasets}. There is a wide variety of SFs in the binding affinity prediction task. In general, one can classify them into four groups \cite{LiuJie:2014}: a) Force-field based or physical based SFs; b) Empirical or linear regression based SFs; c) Potential of the mean force (PMF) or knowledge-based SFs; and d) Machine learning based SFs. The present method falls into the last category.

\begin{table}[!ht]
    \centering
    \caption{Summary  of PDBbind datasets used in the present work}
    \begin{tabular}{|l|c|c|}
        \hline
        & Training set complexes & Test set complexes\\ \hline
        CASF-2007 benchmark   & 1105 & 195 \\\hline
        CASF-2013 benchmark   & 3516 & 195 \\\hline
        CASF-2016 benchmark   & 3767 & 290 \\
        \hline
    \end{tabular}
    \label{tab:PDBbind_datasets}
\end{table}

\subsection{Hyperparameters optimization}
\begin{table}[!ht]
    \centering
    \caption{The ranges of model hyperparameters for CV procedure}
    \begin{tabular}{|c|c|}
        \hline
        Parameter & Domain\\ \hline
        $\tau$  & $\{0.5, 1.0, \dots, 6\}$ \\\hline
        $\beta$ & $\{0.5, 1.0, \dots, 6\} \cup \{10, 15, 20\}$\\\hline
        $\mathcal{M}$ & $\{{\rm Adj, Lap, Inv}\}$\\
        \hline
    \end{tabular}
    \label{tab:PDBbind_parameter_domain}
\end{table}
As a rule of thumb, the machine learning models achieve the best performance when their essential parameters are properly optimized. To this end, a 5-fold cross-validation (CV) is carried out to tune the kernel hyperparameters $\mathcal{M}$, $\Omega$, $\beta$, and $\tau$ in the proposed model ${\rm AGL}^{\mathcal{M}}_{\Omega, \beta, \tau}$. For simplicity, we perform the kernel parameter optimization on CASF-2007 benchmark's training data ($N=1105$), the smallest training set among three benchmarks. Based on our previous work \cite{KLXia:2015f, DDNguyen:2017d}, the reasonable range of the hyperparameters are listed in Table \ref{tab:PDBbind_parameter_domain}. Specifically, the scale factor $\tau$ and power parameters $\beta=\kappa \text{~or~} \nu$ are chosen in [0.5, 6] with an increment of 0.5 so that our model can effectively represent the interactions between protein and ligand in a complex. In addition, a high power values such as $\beta$ $\in\{10, 15, 20\}$ is also taken into account to approximate the ideal low-pass filter (ILF) impact \cite{KLXia:2015f}. There are 40 element interactive pairs formed by 4 commonly atom types, \{${\rm C, N, O, S}$\}, in protein and 10 commonly atom types, \{${\rm H, C, N, O, F, P, S, Cl, Br, I}$\}, in ligands. For the generated adjacent and Laplacian matrices, we only consider its positive eigenvalues. From the resulting set of eigenvalues or corresponding atomic descriptors, one can compute 9 descriptive statistical values, namely the sum, minimum, maximum, mean, standard deviation, variance, sums of zeroth, first, and second powers. That gives rise to a total of 390 features.

For a given type of interaction matrix type $\mathcal{M}$ and a given kernel type $\Omega$, we carry out 5-fold on the training data of CASF-2007 to seek the optimal parameters $\beta$ and $\tau$ based on the averaged Pearson correlation coefficient value $(R_p)$. Fig. S1 in the supplement material  reports the best models with associated $R_p$ in this experiment. The optimal models are  (${\rm AGL}^{\rm Adj}_{{\rm E}, 6, 2.5}$, $R_p = 0.748$), (${\rm AGL}^{\rm Lap}_{{\rm E}, 10, 3.5}$, $R_p = 0.74$), (${\rm AGL}^{\rm Inv}_{{\rm E}, 1.5, 4.5}$, $R_p = 0.708$), (${\rm AGL}^{\rm Adj}_{{\rm L}, 3.5, 1.5}$, $R_p = 0.749$), (${\rm AGL}^{\rm Lap}_{{\rm L}, 15, 3}$, $R_p = 0.740$), and (${\rm AGL}^{\rm Inv}_{{\rm L}, 3.5, 4}$, $R_p = 0.706$). Among them, ${\rm AGL}^{\rm Adj}_{{\rm L}, 3.5, 1.5}$ is the best model and ${\rm AGL}^{\rm Inv}_{{\rm L}, 3.5, 4}$, $R_p = 0.706$ is the worst one. This finding is no surprise. In fact, adjacent matrix is the simplest one but still effectively captures all the interactions between protein and ligand atoms. Since the GNM-style matrix, i.e. $\mathcal{M}={\rm Inv}$, involves the Moore-Penrose inverse, it admits errors in  eigenvalue calculations.

It is reported in the literature that the multi-scale information can boost the predictor's performance \cite{Opron:2015a,DDNguyen:2017d}. Thus, on top of the optimal one-scale model, we impose another kernel with a different parametrization. We also carry out a similar grid-search procedure as did for the single-scale model to explore  optimized parameters. Based on Fig. S2 in the supplement materials, the best two-kernel models are found at the following
  (${\rm AGL}^{\rm Adj}_{{\rm E},6,2.5; {\rm E}, 4, 2}$, $R_p = 0.75$),
  (${\rm AGL}^{\rm Lap}_{{\rm E},10,3.5; {\rm E}, 5, 1.5}$, $R_p = 0.745$),
  (${\rm AGL}^{\rm Inv}_{{\rm E},1.5,4.5; {\rm E}, 4.5, 2}$, $R_p = 0.714$),
  (${\rm AGL}^{\rm Adj}_{{\rm L},3.5,1.5; {\rm L}, 15, 0.5}$, $R_p = 0.751$),
  (${\rm AGL}^{\rm Lap}_{{\rm L},15,3; {\rm L}, 6, 1}$, $R_p = 0.745$),
  and (${\rm AGL}^{\rm Inv}_{{\rm L},3.5,4; {\rm L}, 10, 1}$, $R_p = 0.715$).
It is clear  that models involving the adjacent matrix, i.e., ${\rm AGL}^{\rm Adj}_{{\rm E},6,2.5; {\rm E}, 4, 2}$ and ${\rm AGL}^{\rm Adj}_{{\rm L},3.5,1.5; {\rm L}, 15, 0.5}$,  still outperform the rest. Finally, we form a consensus model named AGL-Score that is defined by  the mean of the predicted values produced by those two aforementioned AGL models. That consensus model is utilized for the predictions of all benchmarks in this work.

\subsection{Performance and discussion}

\begin{figure}[!htb]
    \centering
    \includegraphics[width=0.6\textwidth]{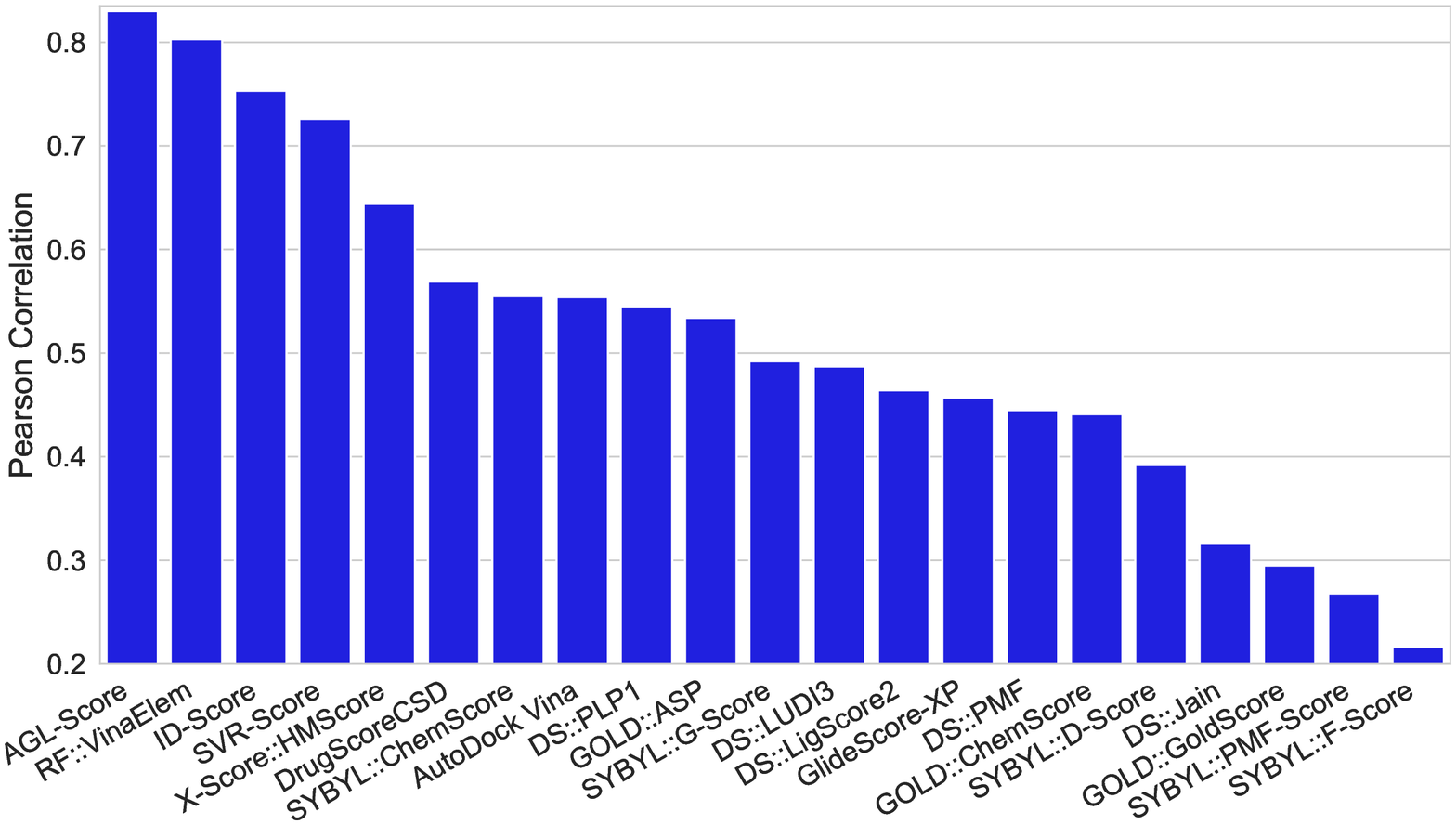}
    \caption{Performance comparison of different scoring functions on CASF-2007 benchmark. The Pearson correlation coefficients of other methods are taken from Refs. \cite{RenxiaoWang:2009Compare,Pedro:2010Binding, IDScore:2013,HLi:2015}. The proposed algebraic graph learning based scoring function, AGL-Score, achieves $R_p=0.83$ and RMSE=1.864 kcal/mol.}
    \label{fig:v2007-benchmark}
    \includegraphics[width=0.6\textwidth]{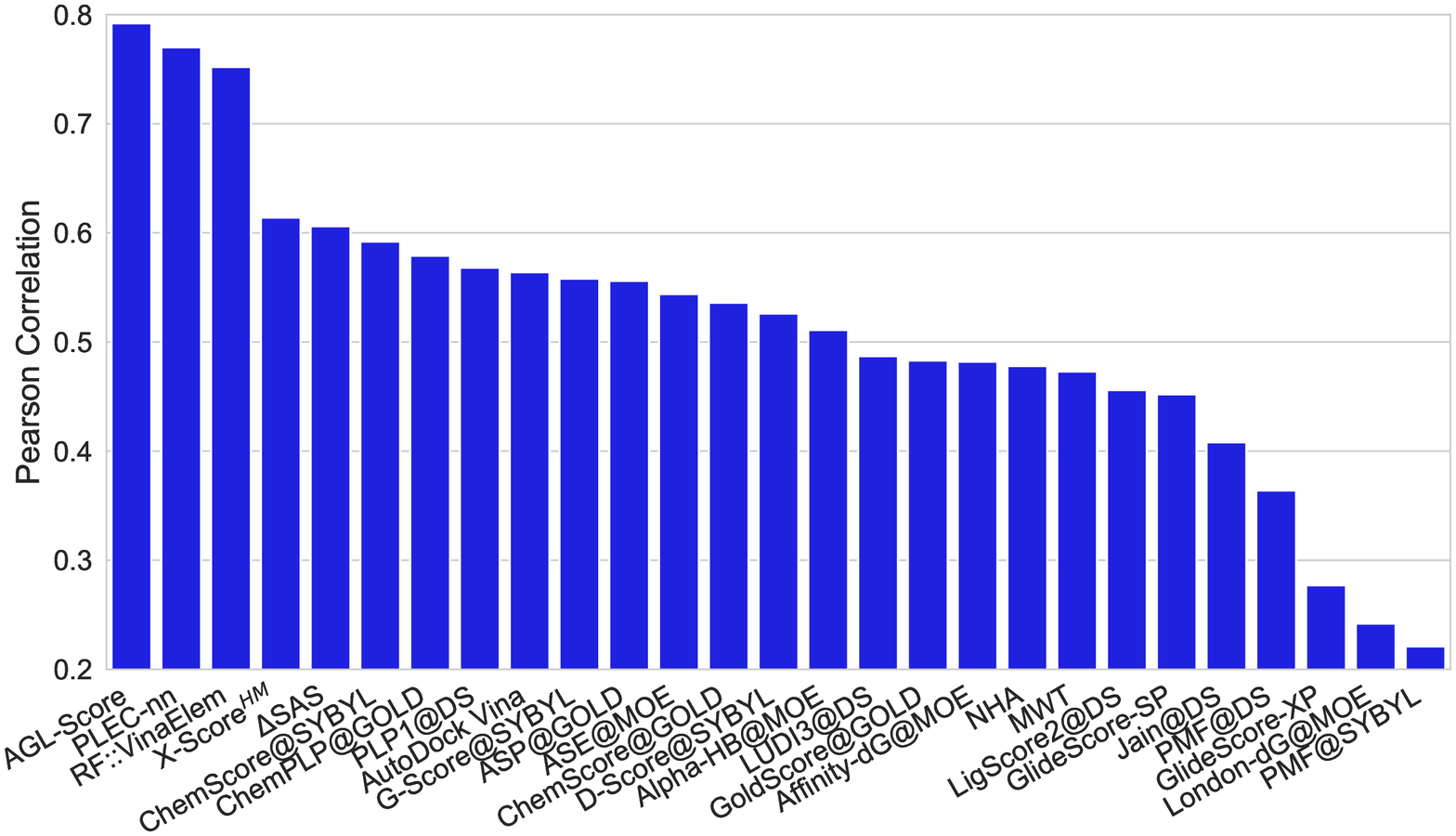}
    \caption{Performance comparison of  different scoring functions on CASF-2013 benchmark. The Pearson correlation coefficients of other methods are taken from Refs. \cite{YLi:2014}, \cite{HongJianLi:2015} and \cite{wojcikowski2018development}. The proposed algebraic graph learning based scoring function, AGL-Score, achieves $R_p=0.792$ and ${\rm RMSE}=1.973$ kcal/mol.}
    \label{fig:v2013-benchmark}
    \includegraphics[width=0.6\textwidth]{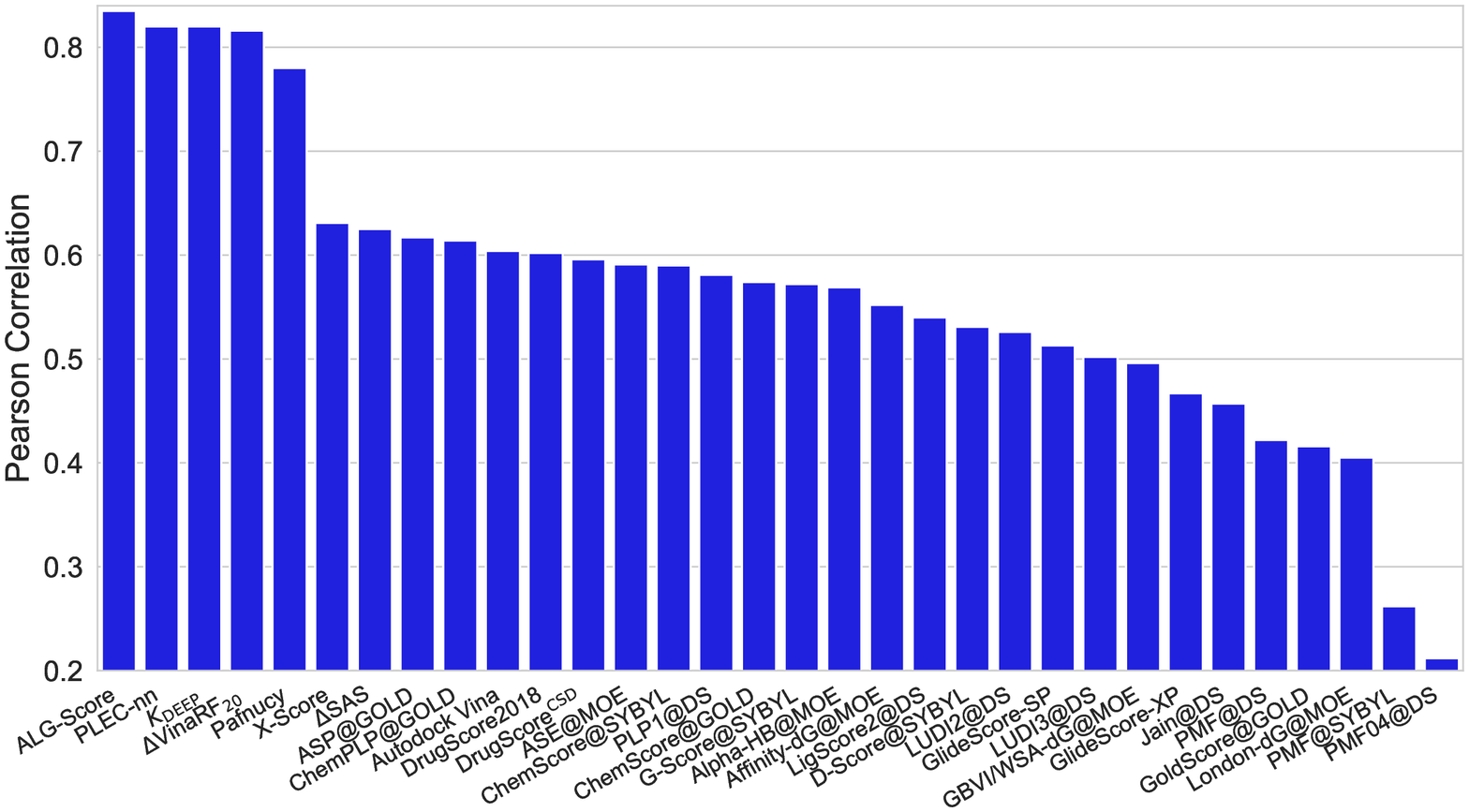}
    \caption{Performance comparison of different scoring functions on CASF-2016. The Pearson correlation coefficients of other methods are taken from Refs. \cite{su2018comparative, jimenez2018k, stepniewska2018development, wojcikowski2018development}. The proposed algebraic graph learning based scoring function, AGL-Score, achieves $R_p=0.835$ and ${\rm RMSE}=1.732$ kcal/mol.}
    \label{fig:v2016-benchmark}
\end{figure}

\begin{figure}[!htb]
    \centering
    \includegraphics[width=0.8\textwidth]{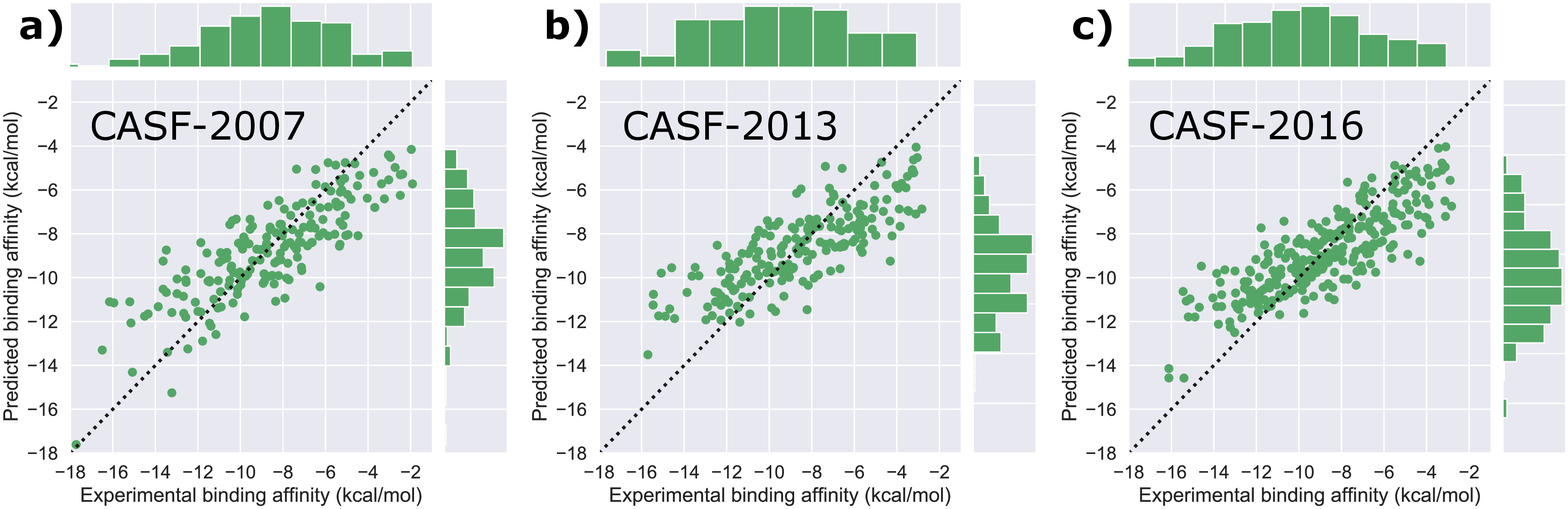}

     \caption{Correlation plot between AGL-Score predictions and experimental data various benchmarks. (a) CASF-2007: Pearson correlation coefficient $R_p=0.83$ and ${\rm RMSE}=1.864$ kcal/mol; (b) CASF-2013: Pearson correlation coefficient $R_p=0.792$ and ${\rm RMSE}=1.973$ kcal/mol; (c) CASF-2016: Pearson correlation coefficient $R_p=0.835$ and ${\rm RMSE}=1.732$ kcal/mol.}
    \label{fig:pred-exp-casf}
\end{figure}

First of all,  we validate the scoring power of our proposed AGL-Score using the CASF-2007 benchmark. We train two AGL models, namely ${\rm AGL}^{\rm Adj}_{{\rm E},6,2.5; {\rm E}, 4, 2}$ and ${\rm AGL}^{\rm Adj}_{{\rm L},3.5,1.5; {\rm L}, 15, 0.5}$ on the refined set ($N=1105$) of the PDBbind v2007 excluding the test set ($N=195$) of CASF-2007 benchmark. For the prediction task, we carry out each AGL model up to 50 times. The average of all the predicted values is used as the predicted binding affinity of the AGL model. It is noted that the energy unit in the PDBbind database is pKd. For the kcal/mol unit conversion, we multiply the predicted values by -1.3633.
In addition, we are interested in comparing the predictive power of our AGL-Score with various state-of-the-art scoring functions introduced in the Literature. \cite{RenxiaoWang:2009Compare,Pedro:2010Binding, IDScore:2013,HLi:2015}. Fig. \ref{fig:v2007-benchmark} illustrates such comparison. Clearly, our proposed model is the most accurate scoring function in this benchmark with Pearson correlation coefficient value $R_p=0.830$ and ${\rm RMSE}=1.864$ kcal/mol. The runner-up is RF::VinaElem with reported $R_p=0.803$ \cite{HLi:2015}. This comparison confirms the scoring power of our proposed model. Furthermore, the correlation visualization between our predicted values and the experimental data is depicted in Fig. \ref{fig:pred-exp-casf}a.

In the second benchmark, i.e. CASF-2013, its training data ($N=3516$) is compiled based on refined set of PDBbind v2015 excluding its test set ($N=195$).
we carry out a similar prediction procedure as of the previous one. Interestingly, the proposed model again outperforms the state-of-the-art scoring functions adopted from Refs. \cite{YLi:2014, HongJianLi:2015} as seen in Fig. \ref{fig:v2013-benchmark}. Specifically, our AGL-Score attains Pearson correlation coefficient value $R_p=0.792$ and ${\rm RMSE}=1.973$ kcal/mol followed by PLEC-nn model with $R_p=0.77$ \cite{wojcikowski2018development}. In addition, Fig. \ref{fig:pred-exp-casf}b provides a scatter plot to illustrate the correlation between our predicted values and experimental results.

CASF-2016 is the last benchmark considered in this work. It is also the latest CASF released by PDBbind database \cite{su2018comparative}. We train the AGL model based on the refined set of PDBbind v2016 excluding the benchmark's test set. In this experiment, both training data ($N=3767$) and test data ($N=290$) are slightly larger than their predecessor, CASF-2013. There is a number of scoring functions having the reported performance on this benchmark. Specially, $K_{\rm DEEP}$ \cite{jimenez2018k}, Pafnucy \cite{stepniewska2018development}, and PLEC-nn \cite{wojcikowski2018development} scoring functions make use of the deep learning architectures. Also, various types of scoring functions have been conducted by the PDBbind team \cite{su2018comparative}. We compare the proposed AGL-Score to the aforementioned methods in Fig. \ref{fig:v2016-benchmark}. Our AGL-Score is still superior to its counterparts with $R_p=0.835$ and ${\rm RMSE}=1.732$ kcal/mol. The second best approaches in the chart are $K_{\rm DEEP}$ and PLEC-nn with reported $R_p=0.82$. This result confirms the accuracy and reliability of the AGL model on the diversified binding affinity datasets. Finally, the comparison between the predicted affinities of AGL model and the experimental values is depicted in Fig. \ref{fig:pred-exp-casf}c.

\section{Availability}
The algebraic graph learning based scoring function model is implemented on our own hosted server at \url{http://weilab.math.msu.edu/AGL/}. User just needs to provide the ligand 3D structure in sdf or mol2 format and protein 3D structure in pdb format for the binding affinity prediction. To maintain  reliable accuracy, the provided structures must bind to each other. If the calculation needs a longer time than usual, the user can provide email for the job completion notice.

%
%

\section{Conclusion}
Algebraic graph theories are commonly used in the study of molecular and biomolecular systems. However, most algebraic graph theory based models are not as accurate as their counterparts \cite{Pires:2014, LQuan:2016,ZXCang:2017a, ZXCang:2017c}. Motivated by our previous work on the multigraph approaches for B-factor predictions \cite{DBramer:2018a}, we propose a novel algebraic  graph learning based scoring function (AGL-Score) for dealing with drug design related problems. The proposed AGL model makes use of  multiscale weight colored subgraphs to encode the essential physical and biological information,  such as hydrogen bonds, electrostatics, van der Waals interactions, hydrophilicity, hydrophobicity, etc.,  presented in the high-dimension space into the low-dimension representation of molecular and biomolecular structures.  The constructions of three types of subgraphs are discussed in this work, namely adjacent matrix, Laplacian matrix, and pseudo-inverse of Laplacian matrix. The eigenvalues calculated from such matrices are used to characterize the biological and physical features of molecules and biomolecules.  In this work, we mainly focus on the binding affinity datasets to demonstrate the robustness, accuracy, and reliability of the proposed model.  To this end,  three mainstream benchmark tests on scoring power assessment, namely CASF-2007 \cite{RenxiaoWang:2009Compare}, CASF-2013 \cite{YLi:2014}, and CASF-2016 \cite{su2018comparative}, are utilized. The results of benchmark tests reveal the superior performance of the proposed scoring function over the state-of-the-art methods. Extensive numerical experiments rigorously confirm the accuracy and reliability of AGL models on various protein-ligand binding affinity datasets.

In addition to the confirmed accuracy and reliability, another major advantage of the present AGL model is its simplicity. Only raw structural inputs regarding atom types and coordinates are used. There is no need for any molecular force field. Additionally, the present model is robust without invoking to  complicated data processing and optimization procedures.

\section*{Funding}
This work was supported in part by  NSF Grants DMS-1721024 and DMS-1761320, and NIH grant  GM126189. DDN and GWW are also funded by Bristol-Myers Squibb and Pfizer.
\vspace*{-12pt}
\clearpage

\section*{Literature cited}
\renewcommand\refname{}


\end{document}